\documentclass[preprint,prl,aps,amsfonts,amssymb,amsmath,showpacs]{revtex4}
\usepackage{graphicx}% Include figure files
\usepackage{dcolumn}% Align table columns on decimal point

\usepackage{amssymb}
\usepackage{graphicx}
\usepackage{subfigure}
\usepackage{float}

\usepackage{amsmath}
\usepackage{color}
\usepackage{booktabs}
\usepackage{multirow}

\usepackage{tabularx}
\graphicspath{{figures/}}
\setkeys{Gin}{width=\linewidth}

\begin{document}

\title{Proof-of-principle experiment of reference-frame-independent quantum key distribution with phase coding}

\author{Wen-Ye Liang$^1$, Shuang Wang$^1$, Hong-Wei Li$^1$, Zhen-Qiang Yin$^1$$^{*}$,\\ Wei Chen$^1$$^{*}$, Yao Yao$^1$, Jing-Zheng Huang$^1$, Guang-Can Guo$^1$, and Zheng-Fu Han$^1$$^{*}$\\}

\affiliation{ $^1$Key Laboratory of Quantum Information,University of Science and
Technology of China,Hefei 230026,China\\
$^{*}$ Correspondence to: (Z.Q.Y) yinzheqi@mail.ustc.edu.cn, (W.C.) kooky@mail.ustc.edu.cn,\\ (Z.F.H) zfhan@ustc.edu.cn.}
\date{\today}

\begin{abstract}
We have demonstrated a proof-of-principle experiment of reference-frame-independent phase coding quantum key distribution (RFI-QKD) over an 80-km optical fiber. After considering the finite-key bound, we still achieve a distance of 50 km. In this scenario, the phases of the basis states are related by a slowly time-varying transformation. Furthermore, we developed and realized a new decoy state method for RFI-QKD systems with weak coherent sources to counteract the photon-number-splitting attack. With the help of a reference-frame-independent protocol and a Michelson interferometer with Faraday rotator mirrors, our system is rendered immune to the slow phase changes of the interferometer and the polarization disturbances of the channel, making the procedure very robust.\\
\end{abstract}

\maketitle

To ensure the security of sensitive data transmission, a series of keys must be securely transmitted between distant users, referred to here as Alice and Bob. Recently, the quantum key distribution (QKD)\cite{BB84, Gisin02} has become useful for distributing secret keys securely. The use of QKD over fibers and free space has been demonstrated many times\cite{Tokyo11, SwissQ11, SECOQC09}. Currently, there are even commercial QKD systems available\cite{Beijing,Wh1,Wh2}.

In most QKD systems, a shared reference frame between Alice and Bob is required. For example, the alignment of polarization states for polarization encoding QKD or interferometric stability for phase encoding QKD plays an important role in those systems. Although alignment operations have been shown to be feasible, they do require a certain amount of time and cost to perform. As an alternative, Laing et al. proposed a reference-frame-independent (RFI) protocol\cite{Laing10} to eliminate the requirement of alignment. This protocol uses three orthogonal bases ($X$, $Y$ and $Z$), in which the $X$ and $Y$ bases are used to estimate Eve's information, and the $Z$ basis is used to obtain the raw key. The states in the $Z$  basis, such as the time-bin eigen-states, are naturally well-aligned, whereas the states in $X$ and $Y$ are superpositions of the eigen-states in $Z$. RFI-QKD could be very useful in several scenarios, such as earth-to-satellite QKD and path-encoded chip-to-chip QKD \cite{Laing10}. However, real-life RFI-QKD systems are vulnerable to the photon-number-splitting (PNS) attack \cite{PNS1, PNS2, PNS3}because a weak coherent light source is usually used instead of a single-photon source. To our knowledge, there has not yet been an experimental demonstration of RFI-QKD in a long-distance fiber, performed in a way that is secure against a PNS attack\cite{frfi}.

However, in the RFI protocol, we must use a finite number of signals to estimate the optimal secure key rate. If Alice and Bob wait for too long, our result will be bad due to misalignment of the frames. Hence, we must consider this protocol in finite-key scenarios. A method for estimating key rate has been described in \cite{FiniteNJP}.

In this letter, a new data analysis method for decoy states in the RFI-QKD protocol is proposed. We provide an experimental demonstration of RFI-QKD with the decoy method. The secure key bits can be generated by our system with up to a 50-km quantum channel distance in finite-key scenarios.

\part{Results}

\section{Theoretical analysis with decoy states}

\subsection{Review of the protocol}

The encoding in RFI-QKD is very similar to the six states protocol \cite{sixstates}. We denote that $|0\rangle$ and $|1\rangle$ consist of the $Z$ basis, $|+\rangle = (|0\rangle+|1\rangle)/\sqrt{2}$ and $|-\rangle = (|0\rangle-|1\rangle)/\sqrt{2}$ consist of the $X$ basis, $|+i\rangle = (|0\rangle+i|1\rangle)/\sqrt{2}$ and $|-i\rangle = (|0\rangle-i|1\rangle)/\sqrt{2}$ consist of the $Y$ basis. For simplicity, we define $X_{A(B)}$, $Y_{A(B)}$ and $Z_{A(B)}$ as Alice(Bob)'s local measurement frames for the $X$, $Y$ and $Z$ bases respectively. In a QKD experiment with well-aligned measurement frames, Alice and Bob should make sure that $X_A=X_B=\sigma_X$, $Y_A=Y_B=\sigma_Y$, $Z_A=Z_B=\sigma_Z$, in which $\sigma_X$, $\sigma_Y$, and $\sigma_Z$ are Pauli operators. However, meeting this requirement may not be easy. One can imagine that $|0\rangle$ and $|1\rangle$ are time-bin eigen-states, and further assume that the quantum channel or interferometer introduces an unknown and slowly time-varying phase $\beta$ between $|0\rangle$ and $|1\rangle$. This implies the following:
\begin{gather}
Z_A = Z_B,\\
X_B = \cos\beta X_A + \sin\beta Y_A,\\
Y_B = \cos\beta Y_A - \sin\beta X_A.
\end{gather}
In each round, Alice chooses one of the encoding states and sends it to Bob through the quantum channel, and Bob measures the incoming photon with $X_B$, $Y_B$ or $Z_B$, chosen at random. After running the protocol for the appropriate number of rounds $N$, we can calculate the bit error rate for the $Z_AZ_B$ basis:
\begin{gather}\label{Ezz}
E_{ZZ} = \dfrac{1 - \langle Z_AZ_B\rangle}{2}.
\end{gather}
Here, $\beta$ should be nearly constant during the $N$ trials. $C$ is used to estimate Eve's information:
\begin{gather}\label{C}
C = \langle X_AX_B\rangle^2 + \langle X_AY_B\rangle^2 + \langle Y_AX_B\rangle^2 + \langle Y_AY_B\rangle^2.
\end{gather}
In a practical QKD system, usually $E_{ZZ}\leq15.9\%$, so the secret key bit rate is $R = 1 - h(E_{ZZ}) - I_E$, where $h(x)$ is the Shannon entropy function. Eve's information $I_E$ is given by
\begin{gather}\label{IE}
I_E= (1 - E_{ZZ})h(\dfrac{1 + v_{max}}{2}) -  E_{ZZ}h(\dfrac{1 + f(v_{max})}{2}),
\end{gather}
in which,
\begin{gather}
v_{max} = min[\dfrac{1}{1 - E_{ZZ}}\sqrt{C/2},1],\\
f(v_{max}) = \sqrt{C/2 - (1 - E_{ZZ})^2v_{max}^2}/E_{ZZ}.
\end{gather}

\subsection{Decoy states method for the RFI-QKD system}

The results mentioned above are based on the use of a single-photon source. Practical QKD implementations using a weak coherent light source must also use the decoy states method to overcome a PNS attack in a long-distance scenario \cite{HwangPRL03, LoPRL05, WangPRL05}. However, the original decoy states method cannot be applied to the RFI system directly. Here, we discuss how to develop decoy states for RFI-QKD implementations.

Assume that Alice randomly modulates the weak coherent laser pulses with three mean photon numbers $\mu$, $\nu$ ($\mu>\nu$) and $0$, which are called signal, decoy, and vacuum
pulses, respectively. For every intensity, Alice and Bob perform the RFI-QKD protocol, and then they obtain the counting rates $Y_\mu$, $Y_\nu$, and $Y_0$ for signal pulses, decoy pulses and vacuum pulses, respectively. Alice and Bob also obtain the error rates $E_{\mu ZZ}$, $E_{\mu xy}$ and $E_{\nu xy}$, (where $x,y=X,Y$). For example, $E_{\mu XY}$ represents the error rate of key bits generated in the case that Alice prepares signal pulses under the $X$ basis while Bob measures the incoming states with the $Y$ basis. According to decoy theory \cite{LoPRL05}, the secret key bits rate $R$ can be calculated as follow:
\begin{gather}\label{R}
R = -Y_\mu h(E_{\mu ZZ})+\mu e^{-\mu}y^L_1(1-I_E),
\end{gather}
Here, $y^L_1$ is the lower bound of the counting rate of the single-photon pulses, and $I_E$ is Eve's information for sifted key bits. $Y_\mu$ and $E_{\mu ZZ}$ are directly observed in the experiment, and $y^L_1$ is given by the following equation \cite{MaPRA05}:
\begin{gather}\label{YL}
y^L_1 = \frac{-\nu^2e^\mu Y_\mu+\mu^2e^\nu Y_\nu-(\mu^2-\nu^2)Y_0}{\mu(\mu\nu-\nu^2)}.
\end{gather}
The next step is to calculate $I_E$ according to (6) or its upper bound.
The upper bound of $I_E$ is related to $c^L_1$, which is defined as the lower bound of $C$ for the single-photon pulses. The upper bound of $I_E$ also depends on the upper bound of the error rate of the key bits generated by single-photon pulses under the $ZZ$ basis $e^U_{1zz}$. According to decoy theory, the following equality applies:
\begin{gather}\label{EU}
e^U_{1ZZ} = \frac{E_{\mu ZZ}Y_\mu-\frac{1}{2}e^{-\mu}Y_0}{\mu e^{-\mu}y^L_1}.
\end{gather}

The challenge is to estimate $c^L_1$ by using $E_{\mu xy}$ and $E_{\nu xy}$. For simplicity, without loss of generality, we assume that $E_{\mu xy}\geqslant 1/2$ and $E_{\nu xy}\geqslant 1/2$ for all $x,y$ (if not, Bob can simply flip his bits corresponding to the relevant basis $x, y$). There are two ways to calculate $c^L_1$:

1. Using the same method as in the original decoy states, as follows:
\begin{gather}
E_{\mu xy}Y_\mu=\frac{1}{2}e^{-\mu}Y_0+e_{1xy}\mu e^{-\mu}y_1+\sum_{n\geqslant 2}e_{nxy}\frac{\mu^n e^{-\mu}}{n!}y_n,
\end{gather}
Here, $e_{nxy}$ $(x,y=X,Y)$ denotes the error rate for the key bits generated by $n$ photon pulses under the $x,y$ basis, $y_n$ represents the counting rate of $n$ photon states. Assuming that $e_{nxy}=1 (n\geqslant 2)$, we obtain that the lower bound of $e_{1xy}$
\begin{gather}
e^L_{1xy}=1-\frac{(1-E_{\mu xy})Y_\mu-\frac{1}{2}e^{-\mu}Y_0}{\mu e^{-\mu}y^L_1}.
\end{gather}
Next, $c^L_1$ is given by $c^L_1=\alpha+\beta$,
where, $\alpha=(1-2Max(1/2,e^L_{1XX}))^2+(1-2Max(1/2,e^L_{1XY}))^2$, $\beta=(1-2Max(1/2,e^L_{1YX}))^2+(1-2Max(1/2,e^L_{1YY}))^2$. Below, we describe the second way to calculate $c^L_1$.

2. We note that
\begin{gather}
E_{\mu XX}Y_\mu=\frac{1}{2}e^{-\mu}Y_0+e_{1XX}\mu e^{-\mu}y_1+\sum_{n\geqslant 2}e_{nXX}\frac{\mu^n e^{-\mu}}{n!}y_n,
\end{gather}
and,
\begin{gather}
E_{\mu XY}Y_\mu=\frac{1}{2}e^{-\mu}Y_0+e_{1XY}\mu e^{-\mu}y_1+\sum_{n\geqslant 2}e_{nXY}\frac{\mu^n e^{-\mu}}{n!}y_n.
\end{gather}
However, $e_{nXX}$ and $e_{nXY}$ are not independent. We assume that Bob obtains some arbitrary two-dimensional density matrices $\rho_+$ and $\rho_-$ after Alice prepares and sends $|+\rangle$ and $|-\rangle$, respectively, through the quantum channel. As described in Ref. \cite{DIQKD}, Alice and Bob's raw key bits are at first distributed in an unbiased fashion (if not, Alice and Bob can perform some classical randomization operations). Thus, it is not restrictive to assume that Eve symmetrizes Alice and Bob's raw key bits, because Eve does not lose any information in this step. Specifically, she can flip Alice and Bob's encoding scheme with a probability of one-half, which is represented as follows:
\begin{gather}
\begin{aligned}
e_{nXX}&=\frac{\langle -|\rho_+|-\rangle+\langle +|\rho_-|+\rangle}{2}.\\
\end{aligned}
\end{gather}
Note that the symmetrization step can also be applied by Alice and Bob in our security analysis. With the help of the Cauchy-Schwarz inequality, we can reformulate the equation:
\begin{gather}
\begin{aligned}
e_{nXY}&=\frac{\langle -i|\rho_+|-i\rangle+\langle +i|\rho_-|+i\rangle}{2}\\
&=\frac{1-Im(\langle +|\rho_+|-\rangle)-Im(\langle -|\rho_-|+\rangle)}{2}\\
&\leqslant \frac{1}{2}+\frac{\big|\langle +|\rho_+|-\rangle\big|+\big|\langle -|\rho_-|+\rangle\big|}{2}\\
&=\frac{1}{2}+\frac{\sqrt{\big|\langle +|\rho_+|-\rangle\big|\big|\langle -|\rho_+|+\rangle\big|}+\sqrt{\big|\langle +|\rho_-|-\rangle\big|\big|\langle -|\rho_-|+\rangle\big|}}{2}\\
&\leqslant \frac{1}{2}+\frac{\sqrt{\big|\langle +|\rho_+|+\rangle\big|\big|\langle -|\rho_+|-\rangle\big|}+\sqrt{\big|\langle +|\rho_-|+\rangle\big|\big|\langle -|\rho_-|-\rangle\big|}}{2}\\
&\leqslant \frac{1}{2}+\sqrt{e_{nXX}(1-e_{nXX})},
\end{aligned}
\end{gather}
Here, $Im(x)$ represents the imaginary part of a real number $x$.
Therefore, we obtain the following:
\begin{gather}
\begin{aligned}
e_{nXX}+e_{nXY}&\leqslant\frac{1}{2}+e_{nXX}+\sqrt{e_{nXX}(1-e_{nXX})}\\
&\leqslant 1.70711.
\end{aligned}
\end{gather}
By adding equations (14) and (15) and applying the above inequality, we find that
\begin{gather}
\begin{aligned}\label{ca}
&e_{1XX}+e_{1XY}\geqslant\\
&1.70711-\frac{(1.70711-E_{\mu XX}-E_{\mu XY})Y_\mu-0.70711e^{-\mu}Y_0}{\mu e^{-\mu}y^L_1}\\
&\triangleq  a.
\end{aligned}
\end{gather}
In the same manner, we find that
\begin{gather}
\begin{aligned}\label{cb}
&e_{1YX}+e_{1YY}\geqslant\\
&1.70711-\frac{(1.70711-E_{\mu YX}-E_{\mu YY})Y_\mu-0.70711e^{-\mu}Y_0}{\mu e^{-\mu}y^L_1}\\
&\triangleq b
\end{aligned}
\end{gather}
With these equations, it is easy to show that $c^L_1=\alpha'+\beta'$,
where, $\alpha'=2(1-a)^2$ and $\beta'=2(1-b)^2$.

Thus, the optimal lower bound of $c_1$ is given by:
\begin{gather}
\begin{aligned}
\label{CL}
c^L_1=Max\{\alpha,\alpha'\}+Max\{\beta,\beta'\}.
\end{aligned}
\end{gather}
This allows us to decide how to evaluate the secure key rate $R$ through the decoy states method: 1. With counting rates $Y_\mu$, $Y_\nu$ and $Y_0$, one can obtain $y_1^L$ by using inequality ~\eqref{YL}. 2. With $y^L_1$ and error rate $E_{\mu ZZ}$, $e_{1ZZ}^U$ is estimated by inequality ~\eqref{EU}. 3. With the error rates $E_{\mu xy}$ ($x,y=X,Y$) and counting rates $y_1^L$, $Y_0$, we obtain $c_1^L$ by using inequality ~\eqref{CL}. 4. We calculate the upper-bound of $I_E$ based on $e_{1ZZ}^U$ and $c_1^L$ using the following equations:

\begin{gather}\label{IE}
I_E= (1 - e_{1ZZ}^U)h(\dfrac{1 + v_{max}}{2}) -  e_{1ZZ}^Uh(\dfrac{1 + f(v_{max})}{2}),
\end{gather}
in which,
\begin{gather}
v_{max} = min[\dfrac{1}{1 - e_{1ZZ}^U}\sqrt{c^L_1/2},1],\\
f(v_{max}) = \sqrt{c^L_1/2 - (1 - e_{1ZZ}^U)^2v_{max}^2}/e_{1ZZ}^U.
\end{gather}
5. Finally, the secure key rate $R$ can be found using equation ~\eqref{R}. This method is applicable to the asymptotic situation. For the finite-key case, we can see that $E_{\mu ZZ}$ and $E_{\mu xy}$ must be modified before we calculate $I_E$.

\subsection{Finite-key bound}

 We use the method for computing the finite-key RIF-QKD bound described in \cite{FiniteNJP}. $p_Z$ is the probability that Alice and Bob choose the $Z$ basis. We assume that the other two bases are chosen with equal probability $p_X = p_Y = p$. As shown previously ~\eqref{C}, there are four measurements needed to estimate $C$, they are $E_{\mu XY}$ ($x,y=X,Y$). For simplicity, and without loss of generality, we assume $E_{\mu xy}\geqslant 1/2$ and $E_{\nu xy}\geqslant 1/2$ for all $x,y$ (if not, Bob can simply flip his bits corresponding to the relevant basis $x,y$).

Experimentally, each value of $E_{\mu xy}$ is estimated using $m = N p^2$ signals. The raw key consists of $n = N p_Z^2$ signals.
As shown previously \cite{FiniteNJP}, under the finite-key scenario, we can correct $E_{\mu ZZ}$ and $E_{\mu xy}$ as $E_{\mu ZZ}'=E_{\mu ZZ}+\delta(n)$ and $E_{\mu xy}'=max\{1/2,E_{\mu xy}-\delta(m)/2\}$, where
\begin{gather}
\delta(k)=\sqrt{\dfrac{\ln(1/\varepsilon_{PE})+2\ln(k+1)}{2k}},
\end{gather}
and $max\{a,b\}$ yields the lesser value of $a$ or $b$.

The key generation rate per pulse against collective attacks is given by \cite{FiniteNJP}:
\begin{gather}
\label{r}
r_{N,col}=-Y_\mu h(E'_{\mu ZZ})+\mu e^{-\mu}y^L_1(1-I_E)-\frac{n}{N}(\frac{1}{n}\log\frac{2}{\varepsilon_{EC}}-\frac{2}{n}\log\frac{1}{\varepsilon_{PA}}-7\sqrt{\dfrac{\log(2/\bar{\varepsilon})}{n}})
\end{gather}

In this article, we set $\varepsilon_{PE}=\varepsilon_{PA}=\varepsilon_{EC}=\bar{\varepsilon}=10^{-5}$. To obtain the correct $I_E$ in the finite-key case, we simply use the method described in the previous section, except that we must adopt $E'_{\mu ZZ}$, $E'_{\mu xy}$ instead of $E_{\mu ZZ}$, $E_{\mu xy}$ as the effective parameters to calculate $I_E$ according to ~\eqref{IE}. Finally, the secure key rate $r_{N,col}$ for the finite-key case can be estimated by ~\eqref{r}.

\section{Experimental setup and results }

The phase coding method was used in our system, and the experimental setup is shown in Fig. 1.

The light pulses generated by Alice's coherent light source are randomly modulated into three intensities of decoy states using an intensity modulator (IM). Then, the quantum states of photons are modulated by a Michelson interferometer with a Faraday rotator mirror (FMI) according to the coding information. Light pulses are attenuated to the single-photon level by a precisely calibrated attenuator before they enter the quantum channel. An SMF-28 single-mode fiber with an attenuation of $0.20dB/km$  is used as a quantum channel between Alice and Bob. To demodulate the information, Bob needs to make measurements of the arriving photons on a randomly and independently selected basis, in which the basis definitions of $X$, $Y$, and $Z$ are the same as those for Alice. There are three possible time-bins of the photons arriving at Bob's single photon detectors (SPD) because there are two FMIs in the system. The SPDs are operating in Geige mode, and their effective gating windows are precisely aligned at the second time-bin.

The FMI used in this system can self-compensate for polarization fluctuations caused by disturbances in the quantum channel \cite{FM}. The quantum states are randomly modulated with the coding of paths and relative phases of photons. In each arm of the FMI, a variable optical attenuator (VOA) acts as the on-off switch to restrain the path of photons, and the relative phases of photons can be controlled by the phase modulator (PM) of the FMI.

In this system, the X, Y and Z bases are chosen to be ($(|0\rangle+e^{i0}|1\rangle)/\sqrt{2}, (|0\rangle+e^{i\pi}|1\rangle)/\sqrt{2}$), ($(|0\rangle+e^{\frac{i\pi}{2}}|1\rangle)/\sqrt{2}, (|0\rangle+e^{\frac{i3\pi}{2}}|1\rangle)/\sqrt{2}$) and ($|0\rangle, |1\rangle$). The coding method for these is as follows: 1) If basis Z is chosen, only one of the two VOAs in Alice's FMI is switched on to allow photons to pass through. Specifically, the time-bin eigen-state $|0\rangle$ or $|1\rangle$ will be determined when Alice switches on the long or the short arm of her FMI, respectively. In this circumstance, Bob can generate his key as long as the detector clicks. That is the code for Alice must be 0 when Bob's code is 1, and vice versa. 2) If basis X or Y is chosen, the two arms of Alice's FMI will be switched on simultaneously, and photons will pass through the two arms with equal probability. The relative phases of the photons can be values from this set: $\{0,\pi/2,\pi,3\pi/2\}$. The values $\{0,\pi\}$ correspond to the $X$ basis, and $\{\pi/2,3\pi/2\}$ correspond to the $Y$ basis.

In Fig. 2, the variation of $\beta$ is random and relatively slow. Every $\beta$ corresponds to a group of QBER values: $E_{\mu xx}$, $E_{\mu xy}$, $E_{\mu yx}$ and $E_{\mu yy}$. We performed counts on 10,000 groups of data, and then plotted the distribution of QBER values in Figure. 3. This figure reveals the random variation in $\beta$ between Alice and Bob, and it also shows our experimental data, measured for the case in which $\beta$ is universally randomly varying .

Fig. 4(a) shows the key generation rate per pulse only for decoy states and compares the rates with those of Fig. 4(b) by using finite-key analysis. In finite-key analysis, being able to calculate the secret key rate by our protocol depends strongly on the number of quantum signals sent in the stationary segment. Hence, the key rate for three different stationary segments is shown in Fig. 5. In the 5-s case, the number of signals is approximately 15,000 at 0 km, and because this number is small, the finite key effect is strong. Using the same experimental parameters and estimation techniques, the key generation rate of our scheme is similar to the expected value under the RIF scheme.
In our experiment, $E_{ZZ}$ was mainly derived from the dark counts of detectors(e.g. approximately 0.0035 at 0 km and 0.016 at 50 km). More detailed data are shown in Table 1 and Table 2.

\part{Discussion}

In summary, we have experimentally demonstrated a phase coding RFI-QKD system that uses the decoy states method. The system can generate secure key bits via an 80-km optical fiber, and it can effectively resist PNS attacks. In addition, when we consider the finite-key bound, we can obtain secure key bits via a 50-km optical fiber. Our system is intrinsically stable in a slowly varying environment without active alignment, and it benefits from the polarization stability of the FMI. With initiatives for practical QKD underway, we believe that this experiment is timely and that it will bring such QKD systems into practical use.

\part{Methods}

{\bf Device description and experimental setup.}
In this experiment, we use a homemade laser that can emit $1449.85nm$ weak coherent pulses with a $700ps$ pulse width and a $0.052nm$ line width. The FMIs in both Alice and Bob's sites have the same arm-length difference $2m$, to ensure that the time slots of the pulses after the FMIs can be separated completely. The circulator of Bob's system cannot only be used to regulate the light path coupled to one of two SPDs, but it can also be used to resist Trojan horse attacks. The intensities of the signal, decoy, and vacuum states are $\mu = 0.6$, $\nu = 0.2$ and $0$, respectively, and the pulse number ratio is 6:2:1. The single-photon avalanche detectors in our experiment are the id200 model of id Quantique. The dark count probabilities of the detectors, after-pulse probability and detection efficiency, are approximately $4 \times 10^{-5}/gate$, $0.358\%$ and $11\%$, respectively.

We use a personal computer (PC) to control Alice and Bob simultaneously. The entire system is synchronized at $1MHz$. The major limitation comes from the rising and falling times of the commercially available VOAs, which take approximately $250nm$ to switch from maximum to minimum attenuation. The master clock of the system is generated by a PCI-6602 Data Acquisition (DAQ) card (National Instruments) at Alice's site, and it is distributed to Bob through a DG535 delayer (Stanford Research Systems) for accurate synchronization. A PCI-6602 DAQ Card is used to trigger the laser and another DAQ Card USB-6353. The random numbers used to select the basis and states are generated by a software pseudo-random number generator and then transformed to a hardware control signal by a USB-6353 card. The USB-6353 card also records the single-photon detection events from the SPDs, and the collected raw data are transferred to the PC for basis sifting and post processing.

\section{Acknowledgements}
This work was supported by the National Basic Research Program of China (Grants No. 2011CBA00200 and No. 2011CB921200) and the National Natural Science Foundation of China (Grants No. 60921091, No. 61101137, and No. 61201239). Correspondence should be addressed to yinzheqi@mail.ustc.edu.cn,
kooky@mail.ustc.edu.cn and zfhan@ustc.edu.cn.

\section{Author contributions}
For this publication, W.L., S.W. and J.H. constructed the system, performed all the measurements, and analyzed the data.
Z.Y. and H.L. wrote the main manuscript text and W.L. prepared figures 1-4. Y.Y., Z.H. and G.G. provided essential comments to the manuscript.
W.C. and Z.Y. designed the study.
All authors reviewed the manuscript.
The first three authors contributed equally to this letter.

\section{Additional information}
Competing financial interests: The authors declare no competing financial interests.
License: This work is licensed under a Creative Commons license.

PACS numbers: \{03.67.Dd,03.67.Hk\}

\newpage

\begin{table}[h!]
\caption{. Detailed experimental results for Fig. 4(a). ($R_D$ is calculated with decoy states but without the finite key effect. $SEM_{R_D}$ is the standard error of the mean of the secure key generation rate per pulse $R_D$.)}
\arrayrulewidth=1pt
\begin{tabular}{@{}lrrrrrrr@{}}\hline
L(km)          &     0                  &   25                   &           50           &       65               &          75            &        80              & \quad\quad\quad  85\\\hline
$R_D$          & $5.474 \times 10^{-3}$ & $1.468 \times 10^{-3}$ & $3.238 \times 10^{-4}$ & $9.484 \times 10^{-5}$ & $8.223 \times 10^{-6}$ & $1.117 \times 10^{-6}$ &    0\\
$SEM_{R_D}$    & $5.752 \times 10^{-5}$ & $2.276 \times 10^{-5}$ & $1.186 \times 10^{-6}$ & $8.132 \times 10^{-7}$ & $5.783 \times 10^{-7}$ & $4.592 \times 10^{-7}$ &    0\\\hline
\end{tabular}
\end{table}

\begin{table}[h!]
\caption{. Key generation rate per pulse corresponding to Fig.4(b) and Fig.5.}
\arrayrulewidth=1pt
\begin{tabular}{@{}lrrrrrrr@{}}\hline
L(km)  &     0                  &   25                   &          35            &       45               &          50            &        60              &   65\\\hline
$5s$   & $1.307 \times 10^{-3}$ &    $0$                 & $0$                    & $0$                    & $0$                    & $0$                    &    0\\
$50s$  & $3.869 \times 10^{-3}$ & $6.984 \times 10^{-4}$ & $2.819\times 10^{-4}$  & $2.227 \times 10^{-5}$ & $5.967 \times 10^{-6}$ & $0$                    &    0\\
$200s$ & $4.442 \times 10^{-3}$ & $1.025 \times 10^{-3}$ & $5.054\times 10^{-4}$  & $2.061 \times 10^{-4}$ & $9.175\times 10^{-5}$ & $9.009 \times 10^{-6}$ & $3.276 \times 10^{-7}$\\\hline
\end{tabular}
\end{table}

\begin{figure}
\begin{center}
\includegraphics[width=12 cm]{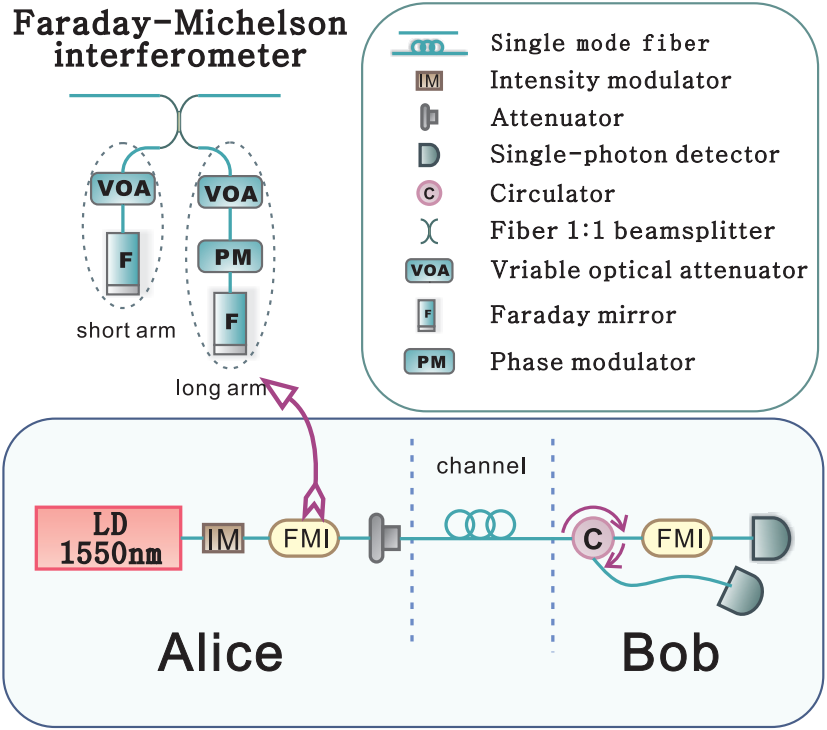}
\caption{ (color online). The experimental setup of the reference-frame-independent quantum key distribution system with decoy states. Channel attenuation is $0.20dB/km$. The arm-length difference of the FMI is $2m$.}
\end{center}
\end{figure}

\begin{figure}
\begin{center}
\includegraphics[width=8 cm]{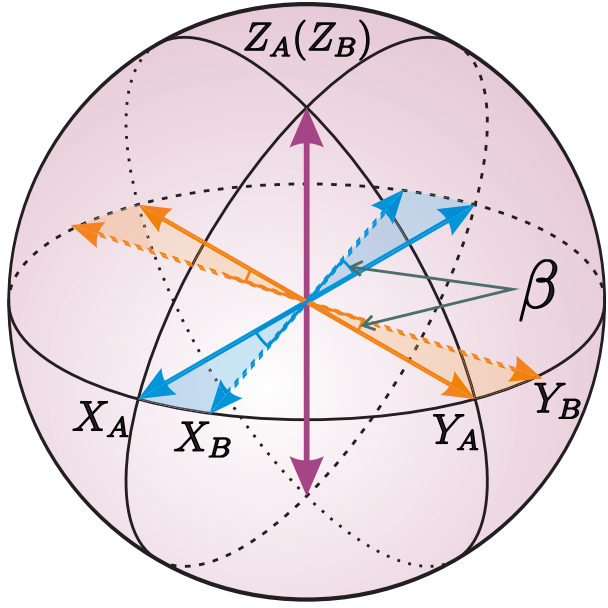}
\caption{ (color online). Three orthogonal states in the phase coding methods. (a) For the X (yellow arrows) and Y (blue arrows) bases, we use $|0\rangle + e^{(i\phi)}|1\rangle$ to express the states. $|0\rangle$ and $|1\rangle$ represent the paths that the pulses travel. $|0\rangle$ is the short arm, $|1\rangle$ is the long arm. $\phi$ is the phase information (b) for the Z (red arrows) basis, which is expressed as $|0\rangle$ or $|1\rangle$. $\beta$ in our system is a time-varying phase between Alice and Bob.}
\end{center}
\end{figure}

\begin{figure}
\begin{center}
\includegraphics[width=12 cm]{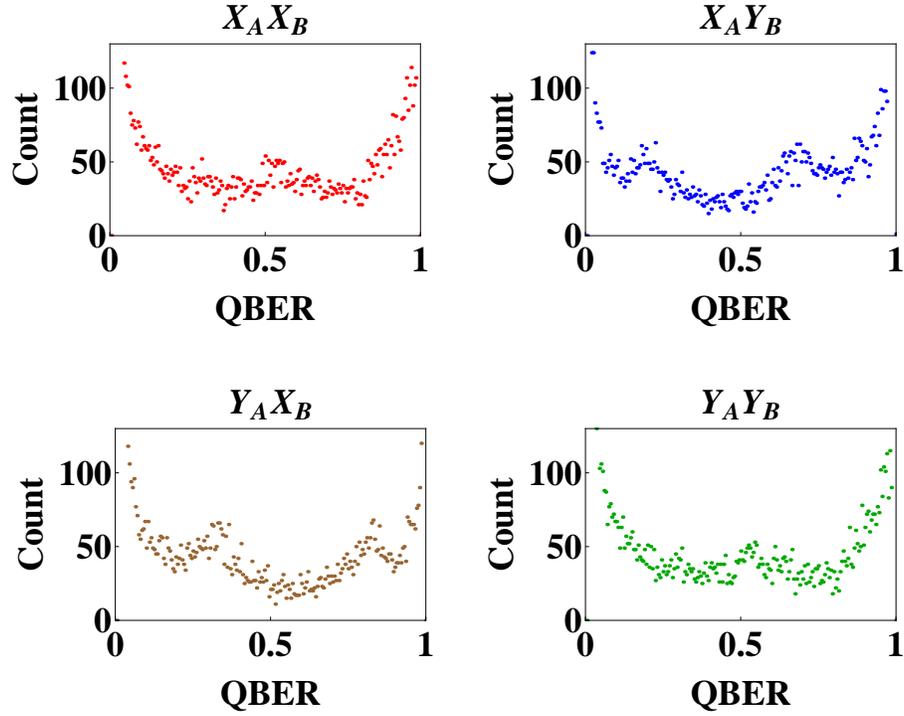}
\caption{ (color online). Distribution of all QBER values in our experiment. QBER values are distributed between 0 and 1. The count of QBER n ($0<n\leq1$) represents the summation of values ranging from $n-0.005$ to $n$.}
\end{center}
\end{figure}

\begin{figure}
\begin{center}
\includegraphics[width=20 cm]{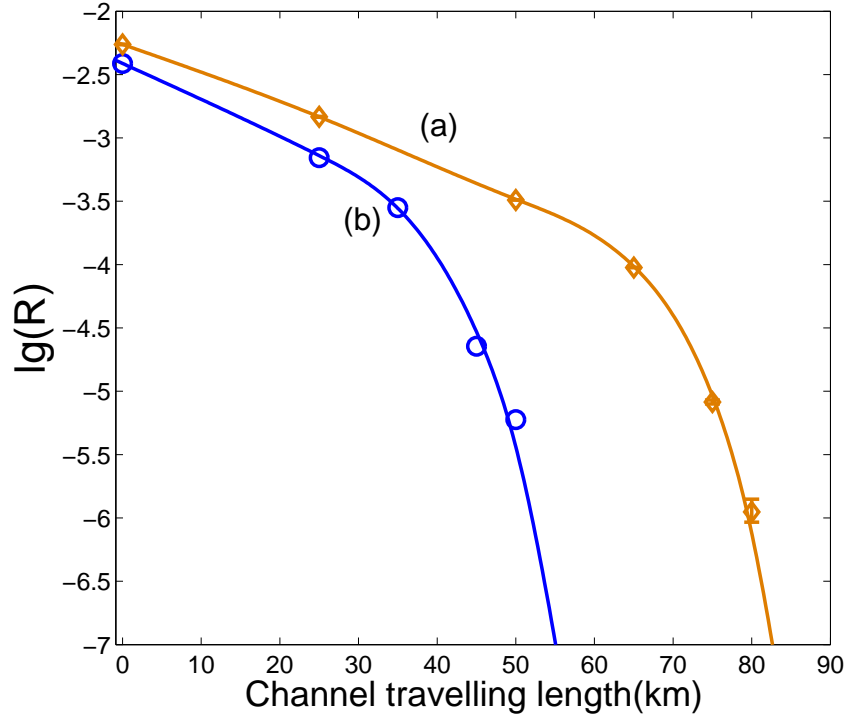}
\caption{ (color online). Calculation (line) and measurement (symbols) of secure key generation rate per pulse with decoy states as a function of channel length. (a) and (b) both use data collected in 50 seconds to calculate the $C$ value. At 0 km, $n \approx m \approx 142,937, E_{\mu zz}\approx0.0035$. (a) Without finite-key analysis, (b) With finite-key analysis.}
\end{center}
\end{figure}

\begin{figure}
\begin{center}
\includegraphics[width=20 cm]{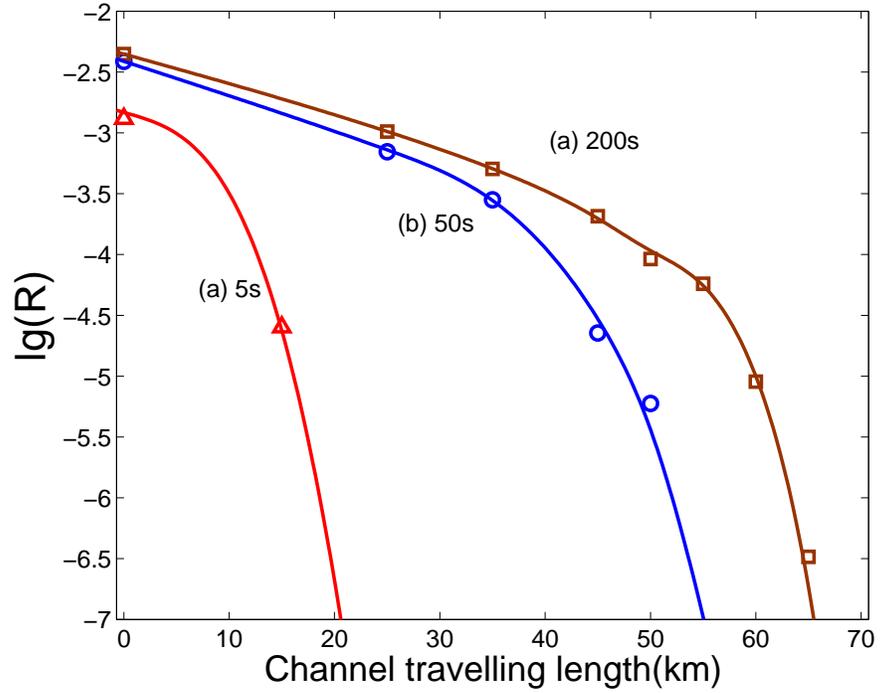}
\caption{ (color online). Calculation (line) and measurement (symbols) of key generation rate per pulse with decoy states for three different numbers of signals. We collected data in different stationary time segments to perform calculations with the same system frequency (from top to bottom: (a) 200 seconds, (b) 50 seconds and (c) 5 seconds).}
\end{center}
\end{figure}

\end{document}